\documentclass[11pt,a4paper]{article}
\usepackage[utf8]{inputenc}
\usepackage[T1]{fontenc}
\usepackage{fixltx2e}
\usepackage{graphicx}
\usepackage{longtable}
\usepackage{float}
\usepackage{wrapfig}
\usepackage{rotating}
\usepackage[normalem]{ulem}
\usepackage{amsmath}
\usepackage{textcomp}
\usepackage{marvosym}
\usepackage{wasysym}
\usepackage{amssymb}
\usepackage{hyperref}
\usepackage{xr}
\usepackage{hyperref}
\usepackage[margin=1in]{geometry}
\usepackage{sidecap}
\usepackage[table]{xcolor}
\hypersetup{
colorlinks,%
citecolor=black,%
filecolor=black,%
linkcolor=blue,%
urlcolor=black
}
\usepackage{setspace}
\usepackage{mathtools}
\expandafter\def\expandafter\quote\expandafter{\quote\small\singlespacing}
\mathtoolsset{showonlyrefs}
\newcommand\mybibliostyle{plainnat}
\usepackage[round, semicolon, authoryear, compress]{natbib}
\date{}
\title{Chromosome-scale shotgun assembly using an \emph{in vitro} method for long-range linkage}
\hypersetup{
  pdfkeywords={},
  pdfsubject={},
  pdfcreator={Emacs 24.3.1 (Org mode 8.2.3c)}}
\begin{document}

\maketitle

Nicholas H. Putnam$^{\text{1*}}$, Brendan O’Connell$^{\text{1,2*}}$, Jonathan C. Stites$^{\text{1}}$, Brandon J. Rice$^{\text{1}}$, Andrew Fields$^{\text{1}}$, Paul D. Hartley$^{\text{1}}$, Charles W. Sugnet$^{\text{1}}$, David Haussler$^{\text{3}}$, Daniel S. Rokhsar$^{\text{4}}$, and Richard E. Green$^{\text{1,2}}$

\begin{quote}
 [1] Dovetail Genomics, 2161 Delaware Ave., Suite A2, Santa Cruz CA 95060; [2] Department of Biomolecular Engineering, University of California, Santa Cruz; [3] UC Santa Cruz Genomics Institute and Howard Hughes Medical Institute, University of California, Santa Cruz; [4] Department of Molecular and Cell Biology, University of California, Berkeley and Department of Energy Joint Genome Institute, Walnut Creek CA; * Equally contributing authors.  
\end{quote}

\begin{abstract}
Long-range and highly accurate \emph{de novo} assembly from short-read data is one of the most pressing challenges in genomics. Recently, it has been shown that read pairs generated by proximity ligation of DNA in chromatin of living tissue can address this problem. These data dramatically increase the scaffold contiguity of assemblies and provide haplotype phasing information. Here, we describe a simpler approach (“Chicago”) based on \emph{in vitro} reconstituted chromatin.  We generated two Chicago datasets with human DNA and used a new software pipeline (“HiRise”) to construct  a highly accurate \emph{de novo} assembly and scaffolding of a human genome with scaffold N50 of 30 Mbp. We also demonstrated the utility of Chicago for improving existing assemblies by re-assembling and scaffolding the genome of the American alligator. With a single library and one lane of Illumina HiSeq sequencing, we increased the scaffold N50 of the American alligator from 508 Kbp to 10 Mbp. Our method uses established molecular biology procedures and can be used to analyze any genome, as it requires only about 5 micrograms of DNA as the starting material.
\end{abstract}
\newpage

A “holy grail” of genomics is the accurate reconstruction of full-length haplotype-resolved chromosome sequences with low effort and cost.  High-throughput sequencing methods have sparked a revolution in the field of genomics. By generating data from millions of short fragments of DNA at once, the cost of re-sequencing genomes has fallen dramatically, rapidly approaching \$1,000 per human genome \citep{pmid24509734}. Substantial obstacles remain, however, in transforming short read sequences into long, contiguous genomic assemblies.

Currently accessible and affordable high-throughput sequencing methods are best suited to the characterization of short-range sequence contiguity and genomic variation.  Achieving long-range linkage and haplotype phasing requires either the ability to directly and accurately read long (i.e., tens of kilobase) sequences, or the capture of linkage and phase relationships through paired or grouped sequence reads.

A number of methods for increasing the contiguity and accuracy of \emph{de novo} assemblies have recently been developed. Broadly, they either attempt to increase the read lengths generated from sequencing or increase the insert size between paired short reads. For example, the PacBio RS II can produce raw reads up to 23 Kbp (median 2 Kbp) in length but suffers from error rates as high as \textasciitilde{}15\% and remains \textasciitilde{}100-fold more expensive than high-throughput short reads \citep{pmid22750884,pmid22827831}. Commercially-available long-reads from Oxford Nanopore are promising but have even higher error rates and lower throughput\citep{goodwin2015}. Illumina’s TruSeq Synthetic Long-Read technology (formerly Moleculo) is limited to 10 Kbp reads maximum \citep{pmid23840927}. CPT-seq is somewhat similar in approach but does not rely on long-range PCR amplification \citep{pmid25327137,pmid25326703}. Despite a number of improvements, fosmid library creation \citep{pmid22800726,lucigen}, remains time-consuming and expensive. To date, the community has not settled on a consistently superior technology for large inserts or long reads that is available at the scale and cost needed for large-scale projects like the sequencing of thousands of vertebrate species\citep{pmid19892720} or hundreds of thousands of humans \citep{pmid24200819}.

The challenge of creating reference-quality assemblies from low-cost sequence data is evident in the comparison of the quality of assemblies generated with today's technologies and the human reference assembly \citep{pmid21102452}. Many techniques including BAC clone sequencing, physical maps, and Sanger sequencing were used to create the high quality and highly contiguous human reference standard with an 38.5 Mbp N50 length and error rate of 1 per 100,000 bases \citep{pmid15496913}. In contrast a recent comparison of the performance of whole genome shotgun (WGS) assembly software pipelines, each run by their developers on very high coverage data sets from libraries with multiple insert sizes, produced assemblies with N50 scaffold length ranging up to 4.5 Mbp on a fish genome and 4.0 Mbp on a snake genome \citep{pmid23870653}.

High coverage of sequence with short reads is rarely enough to attain a high-quality and highly contiguous assembly. This is due primarily to repetitive content on both large and small scales, including the repetitive structure near centromeres and telomeres, large paralogous gene families like zinc finger genes, and the distribution of interspersed nuclear elements such as LINEs and SINEs. Such difficult-to-assemble content composes large portions of many eukaryotic genomes, \emph{e.g.} 60-70\% of the human genome \citep{pmid22144907}. When such repeats cannot be spanned by the input sequence data, fragmented and incorrect assemblies result. In general, the starting point for \emph{de novo} assembly combines deep coverage (50X-200X minimum), short-range (300-500 bp) paired-end “shotgun” data with intermediate range “mate-pair” libraries with insert sizes between 2 and 8 Kbp, and longer range (35 Kbp) fosmid end pairs \citep{pmid21187386,pmid22147368}.  Even this is not completely adequate.

Recently, high-throughput short-read sequencing has been used to characterize the three-dimensional structure of chromosomes in living cells.  Proximity ligation based methods like Hi-C \citep{pmid19815776} and other chromatin capture-based methods \citep{pmid22198700,pmid22495300} rely on the fact that, after fixation, segments of DNA in close proximity in the nucleus are more likely to be ligated together, and thus sequenced as pairs, than are distant regions. As a result, the number of read pairs between intra-chromosomal regions is a slowly decreasing function of the genomic distance between them.  Several approaches have been developed that exploit this information for the purpose of genome assembly scaffolding and haplotype phasing. \citep{pmid24185095,pmid24185094,pmid24270850,pmid25517223}

While Hi-C and related methods can identify long-range chromatin contacts and other biological features of chromosomes at multi-megabase length scales, the principal signal useful for genome assembly and phasing is derived from DNA-DNA contacts on the scale of tens or hundreds of kilobases.  These contacts arise from the polymer physics of the nucleosome-wound DNA fiber, rather than from chromatin biology.  In fact, the large-scale organization of chromosomes in nuclei provides a confounding signal for assembly since, for example, telomeres of different chromosomes are often associated in cells.

We demonstrate here that DNA linkages up to several hundred kilobases can be produced \emph{in vitro} using reconstituted chromatin rather than living chromosomes as the substrate for the production of proximity ligation libraries.  The resulting libraries share many of the characteristics of Hi-C data that are useful for long-range genome assembly and phasing, including a regular relationship between within-read-pair distance and read count.  Combining this \emph{in vitro} long-range mate-pair library with standard whole genome shotgun and jumping libraries, we generated a \emph{de novo} human genome assembly with long-range accuracy and contiguity comparable to more expensive methods, for a fraction of the cost and effort.  This method, called “Chicago” (Cell-free Hi-C for Assembly and Genome Organization), depends only on the availability of modest amounts of high molecular weight DNA, and is generally applicable to any species.  Here we demonstrate the value of this Chicago data not only for \emph{de novo} genome assembly using human and alligator, but also as an efficient tool for the identification of structural variations and the phasing of heterozygous variants.

\section*{Results}
\label{sec-1}
\subsection*{Libraries and sequencing}
\label{sec-1-1}
\begin{figure}
\caption{ A diagram of a Chicago library generation protocol. A) Chromatin (nucleosomes in blue) is reconstituted \emph{in vitro} upon naked DNA (black strand) B) Chromatin is fixed with formaldehyde (thin, red lines are crosslinks). C) Fixed chromatin is cut with a restriction enzyme, generating free sticky ends (performed on streptavidin-coated beads, not shown) D) Sticky ends are filled in with biotinylated (blue circles) and thiolated (green squares) nucleotides. E) Free blunt ends are ligated (ligations indicated by red asterisks). F) Crosslinks are reversed and proteins removed to yield library fragments. }
\includegraphics[width=40em]{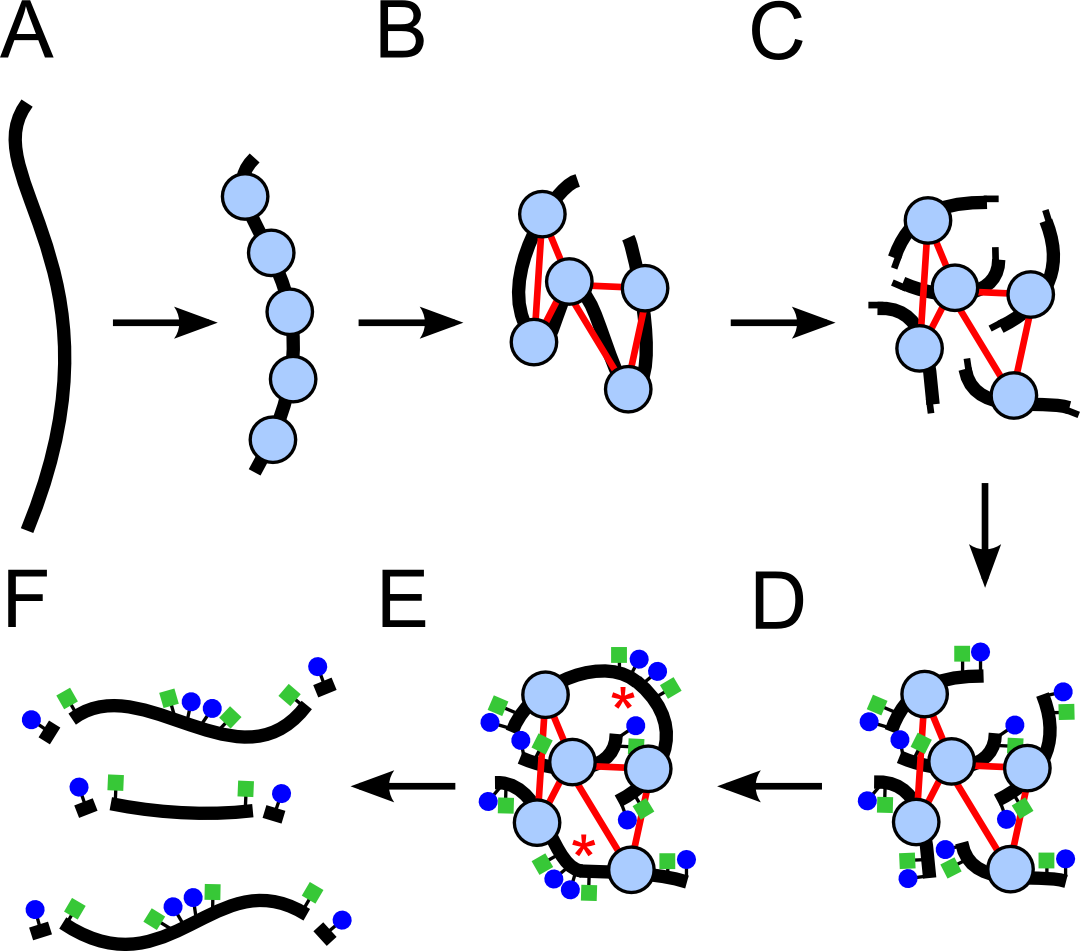}
\label{schematic}
\end{figure}

We extracted 5.5$\mu g$ of high molecular weight DNA for each Chicago library (in fragments of approximately 150 Kbp) from the human cell line GM12878 and from the blood of a wild-caught American alligator.  We reconstituted chromatin by combining the DNA with purified histones and chromatin assembly factors. The reconstituted chromatin was then fixed with formaldehyde and Chicago libraries were generated (Figure \ref{schematic} and Methods).

\begin{figure}

\caption{Read pair separations for several sequencing libraries mapped to hg19. Green: 50 Kbp input human Chicago library. Orange: 150 Kbp input human Chicago library. Light blue: 150 Kbp input human Chicago library. Dark blue: A human Hi-C library\citep{pmid22198700}. Dark vertical lines indicate maximum advertised or demonstrated capabilities for alternative mate-pair technologies. }
\includegraphics[width=40em]{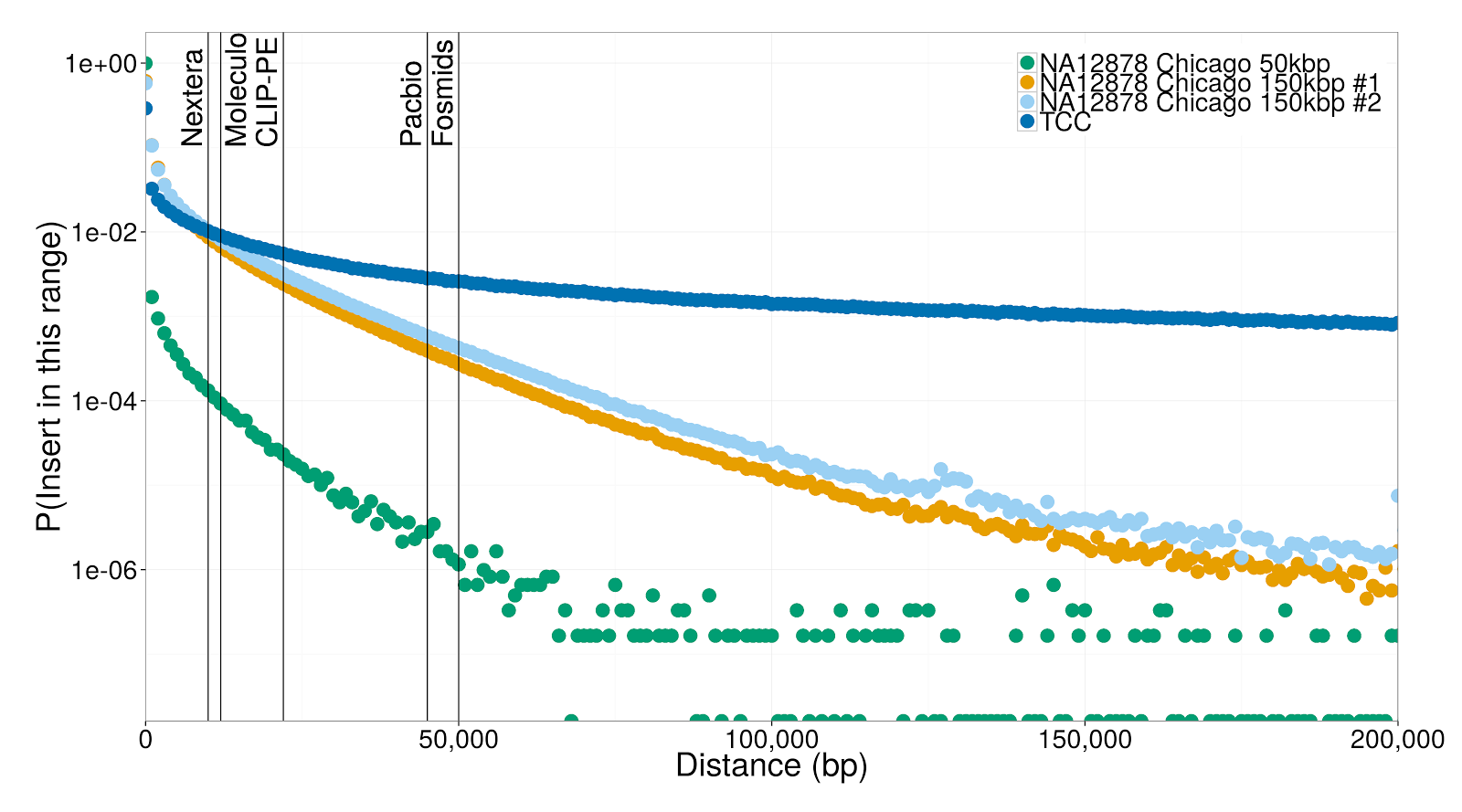}
\label{inserts}
\end{figure}

\begin{figure}

\caption{ Genome coverage (sum of read pair separations divided by estimated genome size) in various read pair separation bins.}
\includegraphics[width=40em]{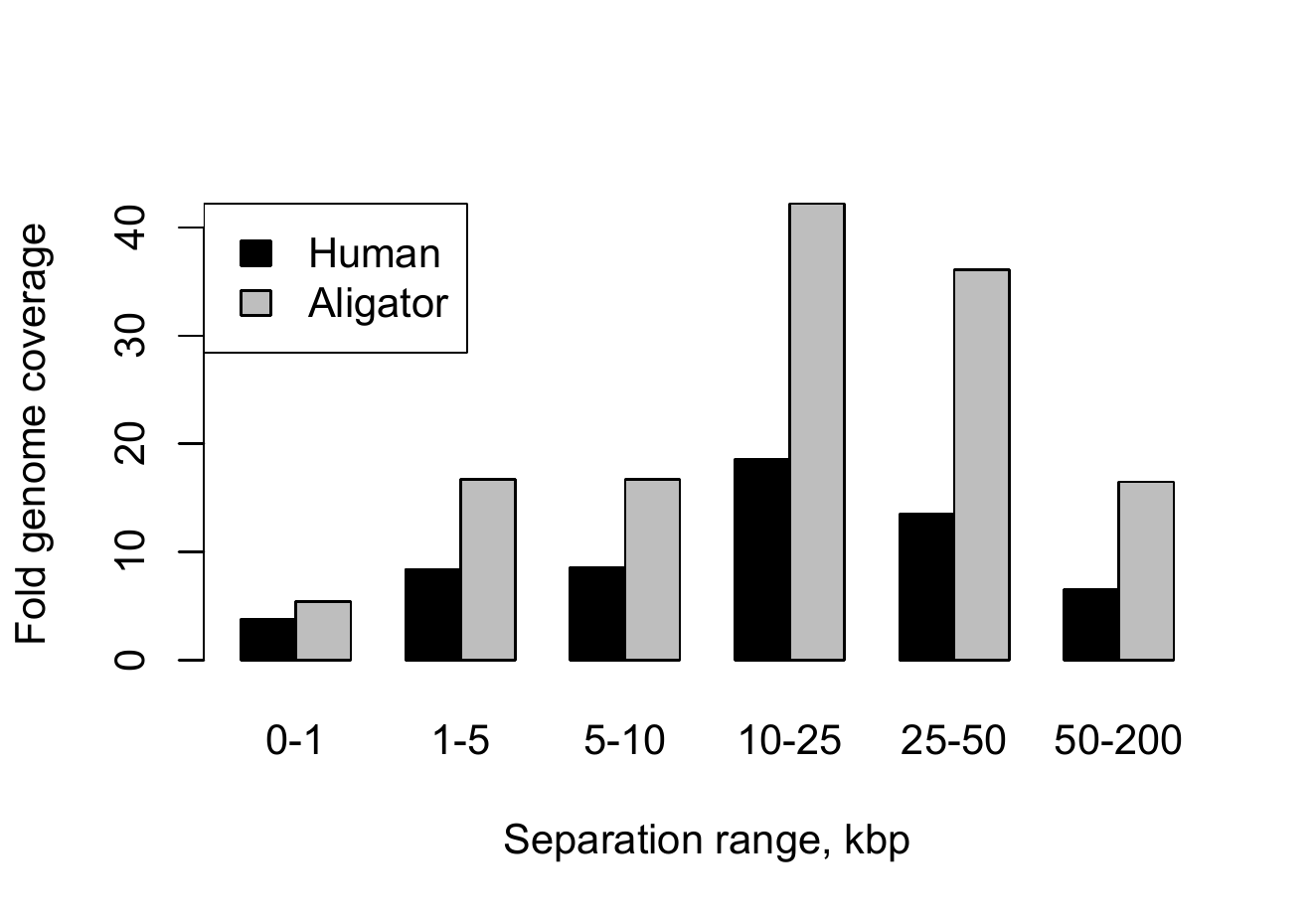}
\label{coverage}
\end{figure}

For the human GM12878 sample we generated two Chicago libraries using the restriction enzymes MboI and MluCI, which generate 4 bp 5’ overhangs. These barcoded libraries were pooled and sequenced on a single Illumina HiSeq 2500 lane in paired 100bp reads, generating 46M MboI and 52M MluCI library read pairs.  For comparison, a third library was prepared from nominally 50 Kbp DNA (Figure \ref{inserts}). For the American alligator (\emph{Alligator mississipiensis}) we constructed a single MboI Chicago library and sequenced it on a single lane, yielding 132M read pairs.
To determine the utility of these data for genome assembly and haplotype phasing, we aligned the GM12878 Chicago data to the reference human assembly (hg19) (Figure \ref{inserts}). The Chicago libraries provided useful linking information up to separations of 150 Kbp on the genome with a background noise rate of approximately one spurious link between unrelated 500 Kbp genomic windows (mean of 0.97 such links). The single lane of sequence from the GM12878 libraries provided linking information equivalent to  3.8X, 8.4X, 8.6X, 18.6X, 13.5X, 6.5X  physical coverage in 0-1, 1-5, 5-10, 10-25, 25-50, and 50-200 Kbp libraries respectively, while for aligator the comparable coverage estimates were 5.4X, 16.7X, 16.7X, 42.2X, 36.1X, 16.5X respectively (Figure \ref{coverage}).

\subsection*{Chicago data for genome scaffolding}
\label{sec-1-2}
To determine the power and utility of data extracted from Chicago libraries, we focused on genome assembly and scaffolding using only generic 300-500 bp insert Illumina shotgun libraries and the Chicago libraries described above.  We also tested the benefit of adding Chicago data to datasets with a broader range of insert sizes.  We initially assembled a previously-described 84x, 101 bp paired-end Illumina shotgun data set from GM12878 \citep{pmid22156294} using MERACULOUS \citep{pmid21876754} into scaffolds with typical (N50) size of  33 Kbp. We mapped the Chicago read pairs to this initial assembly as described in Methods. We found that 68.9\% of read pairs mapped such that both forward and reverse reads had map-quality >= 20 and were thus considered uniquely mapping within the assembly and were not duplicates. 26.8\% of these read pairs had forward and reverse reads that mapped to different contigs and were thus potentially informative for further scaffolding of the assembly.  We also used the same Chicago data to scaffold a Discovar assembly of 50X coverage in paired 250bp reads (\url{ftp://ftp.broadinstitute.org/pub/crd/Discovar/assemblies}) \citep{discovar}.

We developed a likelihood model describing how Chicago libraries sample genomic DNA, and a software pipeline called “HiRise” for breaking and re-scaffolding contigs based on Chicago links (Methods). We compared the completeness, contiguity and correctness at local and global scales of the resulting assembly to assemblies of rich WGS datasets, including extensive coverage in fosmid end pairs created by two of the leading WGS assemblers:  MERACULOUS \citep{pmid21876754} and ALLPATHS-LG (APLG) \citep{pmid21187386} (Table 1). To avoid the arbitrary choices involved in constructing alignment-based comparisons of assembly quality, we based our comparison on the assembled positions of 25.4 million 101 bp sequences that are a randomly-selected subset of all distinct 101 bp sequences that occur exactly once in each haplotype of the diploid NA12878 assembly\citep{pmid21811232}.

\subsection*{Long range scaffolding accuracy}
\label{sec-1-3}
The genomic scaffolds that the HiRise pipeline  produced were longer and had a lower rate of global mis-assemblies than the published MERACULOUS and APLG assemblies, both of which rely on deep coverage in paired fosmid end reads.  Table 1 shows the fraction of the total assembly found in scaffolds containing a misjoin, where a misjoined scaffold is defined as having a stretch of unique 101-mers spanning at least 5 Kbp, 10 Kbp or 50 Kbp from more than one chromosome in the diploid reference.  The table also shows measures of completeness and contiguity for four successive rounds of HiRise assembly compared to other assemblies of NA12878.

\begin{table}[htb]
\caption{\label{tab:scaffolding}Scaffolding results.  Fraction of each assembly in scaffolds containing a misjoin at three different thresholds for identifying misjoins.  Scaffold N50.  50 Kbp separation discrepancy 95\% confidence interval  (95\% CI = x means:  Given a pair of unique 101-mer tags in the assembly, 95\% of them are within 50 Kbp \textpm{} x of each other in the reference.) completeness (\%C); fraction of  101mers misoriented.}
\centering
\begin{tabular}{l|rrr|rlrr|}
 &  & Misjoins &  & N50 & 95\% CI & \%C & \% mis-\\
 & 5 Kbp & 10 Kbp & 50 Kbp & (Mbp) & 50 kbp $\Delta$ &  & oriented\\
\hline
MERACULOUS & 0.032 & 0.030 & 0.022 & 9.1 & 1.3 Kbp & 94.8 & 0.09\\
APLG & 0.245 & 0.187 & 0.130 & 12.1 & 6.4 Kbp & 92.2 & 0.4\\
MERAC (33Kbp N50) + HiRise: & 0.028 & 0.011 & 0.009 & 12.6 & 7.7 Kbp & 94.1 & 1.3\\
MERAC (33Kbp N50) + HiRise0.9.8 & 0.052 & 0.029 & 0.014 & 14.9 & 7.6 Kbp & 94.1 & 1.2\\
Discovar (178Kbp N50) + HiRise: & 0.102 & 0.097 & 0.076 & 29.9 & 3.2 Kbp & 97.9 & 1.4\\
\end{tabular}
\end{table}

Because the DNA ligation events that create Chicago pairs are not constrained to produce read pairs of defined relative strandedness, contig relative orientations during scaffolding must be inferred from read density information.  As a result, the Chicago HiRise scaffolds have a higher rate of mis-oriented 101-mers (1.3\%) than is found in the other assemblies, most occurring in small contigs.  The median size of contigs containing mis-oriented 101-mers was 2.1 Kbp.

\subsection*{Improving the alligator assembly with Chicago data}
\label{sec-1-4}
To further assess the utility of Chicago data for improving existing assemblies, we generated a single Chicago library for the American alligator \emph{Alligator mississippiensis} and sequenced 210.7 million reads on the Illumina HiSeq 2500.  We mapped these data to a \emph{de novo} assembly (N50 81 Kbp) created using publicly available data \citep{pmid25504731} and applied the HiRise scaffolding pipeline.  The resulting assembly had a scaffold N50 of 10.3 Mbp.  To assess the accuracy of these scaffolds, we aligned a collection of 1,485 previously generated  \citep{pmid17307883} bacterial artificial chromosome (BAC) end sequences to the assembly. Of those 1,298 pairs were uniquely aligned by GMAP \citep{pmid15728110} with 90\% coverage and 95\% identity to the genome assembly and the HiRise scaffolded version. In the input assembly, 12.5\% of BAC end pairs were captured in the same scaffold with the expected orientation and separation.  In the HiRise assembly 96.5\% of BAC end pairs were aligned in the same scaffold with 98.1\% of BAC end pairs on same scaffold in correct relative orientation. 5 (0.39\%) BAC end pairs were placed on the same scaffold but at a distance significantly larger than insert size and 14 (1.08\%) were placed on separate scaffolds but far enough from edge of scaffold that distance would be larger than insert size  suggesting a global density of misjoins of less than 1 per 8.36 Mbp of assembly.

\subsection*{Phase accuracy of Chicago read pairs}
\label{sec-1-5}
As shown for Hi-C data \citep{pmid24185094}, read pairs formed by proximity ligation are nearly always the product of ligation between a single contiguous DNA strand. Thus, read pairs where both the forward and reverse read cover a heterozygous site can be used to directly read haplotype phase as is done using fosmid or other mate-pair library data. Because the distance covered in Chicago read pairs can be as great as the size of the input DNA, we assessed the accuracy of phase information and its utility for determining haplotype phase in the GM12878 sample. Importantly, because GM12878 derives from an individual that has been trio-sequenced, gold-standard haplotype phase information is available to check the accuracy of Chicago phasing information.  Read pairs that are haplotype informative and that span between 10 Kbp and 150 Kbp are 99.83\% in agreement with the known haplotype phase for GM12878.

\subsection*{Identification of structural variants}
\label{sec-1-6}
\begin{figure}
\caption{ The mapped locations on the GRCh38 reference sequence of Chicago read pairs are plotted in the vicinity of structural differences between GM12878 and the reference.  Each chicago pair is represented both above and below the diagonal.  Above the diagonal, color indicates map quality score on scale shown; below the diagonal colors indicate the inferred haplotype phase of Chicago pairs based on overlap with a phased SNPs. }
\includegraphics[width=25em]{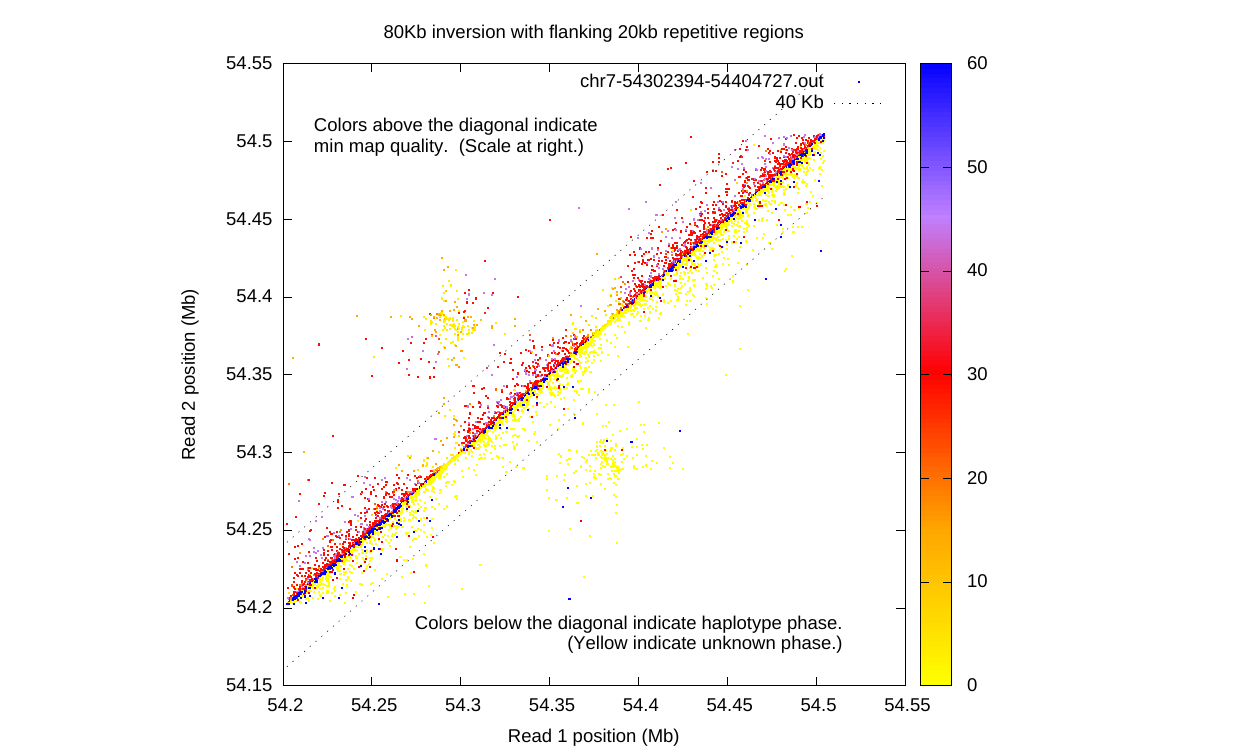}
\includegraphics[width=25em]{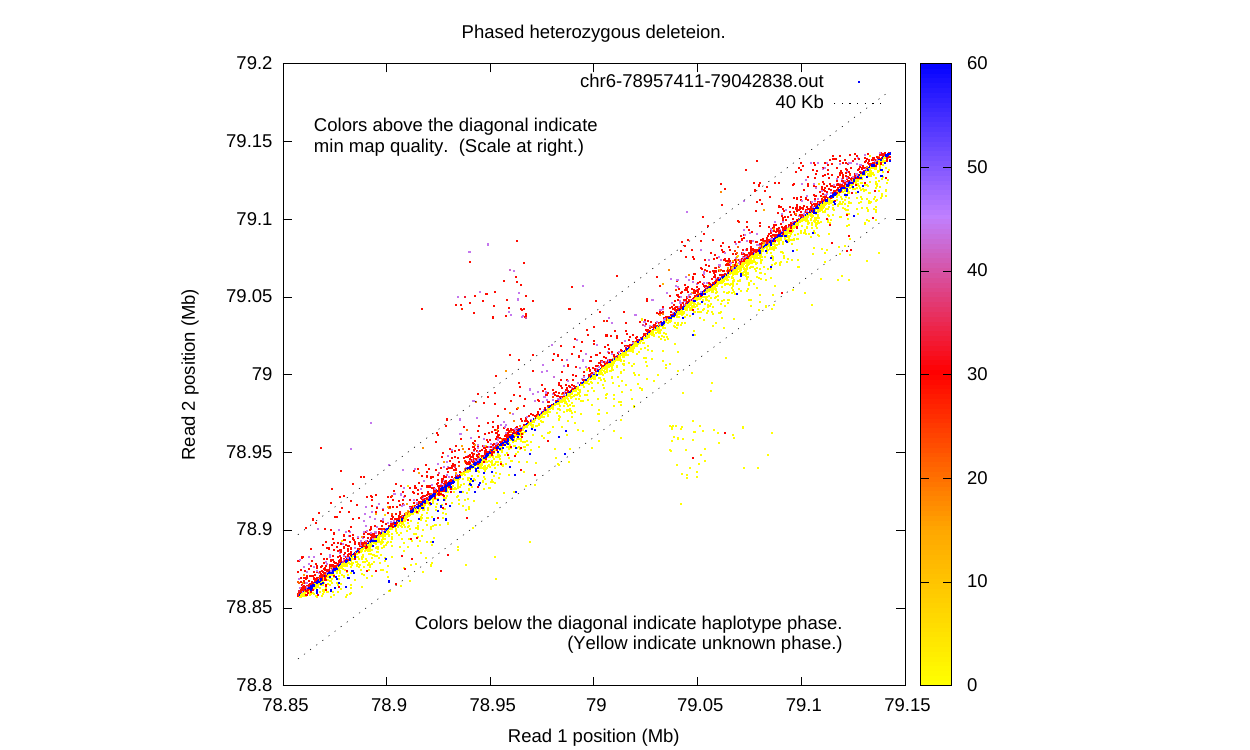}
\label{svarfig}
\end{figure}

Mapping paired sequence reads from one individual against a reference is the most commonly used sequence-based method for identifying differences in genome structure like inversions, deletions and dupications \citep{pmid15895083}.  Figure \ref{svarfig} shows how Chicago read pairs from GM12878 mapped to the human reference genome GRCh38 reveal two such structural differences.  To estimate the sensitivity and specificity of Chicago data for identifying structural differences, we tested a simple maximum likelihood discriminator (Methods) on simulated data sets constructed to simulate the effect of heterozygous inversions.  We constructed the test data by randomly selecting intervals of a defined length L from the mapping of our Chicago NA12878 reads to the GRCh38 reference sequence and assigning each Chicago read pair independently at random to the inverted or reference haplotype, and editing the mapped coordinates accordingly.  Non-allelic homologous recombination is responsible for much of the structural variation observed in human genomes, resulting in  many variation breakpoints that occur in long blocks of repeated sequence\citep{pmid18451855} .    We simulated the effect of varying lengths of repetitive sequence surrounding the inversion breakpoints by removing all reads mapped to within a distance $W$ of them.  In the absence of repetitive sequences at the inversion breakpoints, we found that for 1 Kbp, 2 Kbp and 5 Kbp inversions respectively, the sensitivities (specificities) were 0.76 (0.88), 0.89 (0.89) and 0.97 (0.94) respectively.  Simulating  1 Kbp regions of repetitive (unmappable) sequence at the inversion breakpoints, the sensitivity (specificity) for 5 Kbp inversions was 0.81 (0.76).

\section*{Discussion}
\label{sec-2}
We have described an \emph{in vitro} method for generating long range mate-pair data that can dramatically improve the scaffolding of \emph{de novo} assembled genomes from high-throughput sequencing data. This approach has several advantages over existing methods.
First, Chicago library construction requires no living biological material, i.e., no primary or transformed tissue culture or living organism. The libraries described here were each generated from 5.0 micrograms of input DNA. Furthermore, although the \emph{in vitro} chromatin reconstitution is based on human histones and chromatin assembly factors, DNA from a wide variety of plants, animals, and microbes can be substrate for \emph{in vitro} chromatin assembly using the protocol described (data not shown).
Second, because Chicago data are generated from proximity ligation of chromatin assembled \emph{in vitro} rather than chromatin obtained from \emph{in vivo} sources there is no confounding biological signal (e.g., telomeric clustering, or chromatin looping) to potentially confuse the assembly. Hi-C and or other proximity ligation data generated from \emph{in vivo} chromatin carries within it long-range proximity information that is of biological relevance, but is persistent and potentially confounding for genome assembly. As expected for \emph{in vitro} assembled chromatin, we find a low background rate of noise and a virtual absence of persistent and spurious read pairs.
Third, in contrast to \emph{in vivo} Hi-C methods, the maximum separation of the read pairs  generated is limited only be the molecular weight of the input DNA.  This has allowed us to generate highly contiguous scaffolding of vertebrate genomes using just short fragment Illumina sequence plus Chicago libraries.  To date, high quality scaffolding based on \emph{in vivo} Hi-C libraries have started from assemblies with an order of magnitude more scaffold contiguity than the 30 Kbp N50 input contigs successfully scaffolded by Chicago HiRise.
Fourth, these libraries eliminate the need for creating and sequencing a combination of long range “mate-pair” and fosmid libraries, and do not require the use of expensive, specialized equipment for shearing or size-selecting high molecular weight DNA normally required to create such libraries.
In summary, we have presented a simple DNA library construction and associated bioinformatic methods that generate significantly longer-range genome assembly scaffolds than existing methods. Furthermore, we have demonstrated the usefulness of our data for the discovery of genome variation.
Our methods and results mark a substantial step toward the goal of  accurate reconstruction of full-length haplotype-resolved chromosome sequences with low effort and cost.

\section*{Acknowledgements}
\label{sec-3}

REG is a Searle Scholar and Sloan Fellow. The authors are grateful for the courageous and inspirational assistance of Norma Hadland.

\section*{Data Availability}
\label{sec-4}

For more information and access to the data described here, please see \url{http://www.dovetailgenomics.com/}.

\section*{Methods}
\label{sec-5}
\subsection*{DNA Preparation}
\label{sec-5-1}
DNA was extracted with Qiagen Blood and Cell Midi kits according to the manufacturer’s instructions.  Briefly, cells were lysed, and centrifuged to isolate the nuclei.  The nuclei were further digested with a combination of Proteinase K and RNAse A.  The DNA was bound to a Qiagen genomic column, washed, eluted   and precipitated in isopropanol and pelleted by centrifugation.  After drying, the pellet was resuspended in 200 $\mu L$ TE (Qiagen).

\subsection*{Chromatin assembly}
\label{sec-5-2}
Chromatin was assembled overnight at 27°C from genomic DNA using the Active Motif \emph{in vitro} Chromatin Assembly kit. Following incubation, 10\% of the sample was used for MNase digestion to confirm successful chromatin assembly.

\subsection*{Biotinylation and restriction digestion}
\label{sec-5-3}
Chromatin was biotinylated with iodoacetyl-PEG-2-biotin (IPB).  Following biotinylation, the chromatin was fixed in 1\% formaldehyde at room temperature (RT) for 15 minutes, followed by a quench with 2-fold molar excess of 2.5M Glycine. Excess IPB and cross-linked glycine were removed by dialyzing chromatin in a Slide-A-Lyzer 20 KDa MWCO dialysis cassette (Pierce) against 1L of dialysis buffer (10mm Tris-Cl, pH8.0, 1mM EDTA) at 4°C for a minimum of 3 hours.  Subsequently, the chromatin was digested with either MboI or MluCI in 1X CutSmart for 4 hours at 37°C.  The chromatin was again dialyzed in a 50 KDa MWCO dialysis Flex tube (IBI Scientific \# IB48262) at 4°C for 2 hours, then again with fresh buffer overnight, to remove enzyme as well as short, free DNA fragments.

Dynabead MyOne C1 streptavidin beads were prepared by washing and resuspending in PBS + 0.1\% Tween-20, before adding to chromatin and incubating for 1 hour at RT. The beads were then concentrated on a magnetic concentrator rack, before being washed, re-concentrated, and resuspended in 100 $\mu L$ 1X NEBuffer 2.

\subsection*{dNTP fill-in}
\label{sec-5-4}
To prevent the labeled dNTP's (Figure \ref{schematic}) from being captured during the fill-in reaction, un-bound streptavidin sites were occupied by incubating beads in the presence of free biotin for 15 minutes at RT.  Subsequently, the beads were washed twice before being resuspended in 100 $\mu L$ 1X NEBuffer 2.

Sticky ends were filled in by incubating with dNTPs, including a-S-dGTP and biotinylated dCTP along with 25 U of Klenow (\#M0210M, NEB) in 165 $\mu L$ total volume at 25°C for 40 minutes.  The fill-in reaction was stopped by adding 7 $\mu L$ of 0.5M EDTA.  The beads were then washed twice in Pre-ligation wash buffer (PLWB: 50mM Tris 7.4; 0.4\% Triton X-100; 0.1mM EDTA), before being resuspended in 100$\mu L$ PLWB.

\subsection*{Ligation}
\label{sec-5-5}
Ligation was performed in at least 1mL of T4 ligation buffer at 16°C for a minimum of 4 hours. A large ligation volume was used to minimize cross-ligation between different chromatin aggregates. The ligation reaction was stopped by adding 40$\mu L$ of 0.5M EDTA.  The beads were concentrated and resuspended in 100$\mu L$ extraction buffer (50mM Tris-cl pH 8.0, 1mM EDTA, 0.2\% SDS).  After adding 400ug Proteinase K (\#P8102S, NEB) the beads were incubated overnight at 55°C, followed by a 2 hour digestion with an additional 200 ug Proteinase K at 55°C. DNA was recovered with either SPRI beads at a 2:1 ratio, a column purification kit, or with a phenol:chloroform extraction. DNA was eluted into Low TE (10 mM Tris-Cl pH 8.0, 0.5 mM EDTA).

\subsection*{Exonuclease digestion}
\label{sec-5-6}
DNA was  next digested for 40 minutes at 37°C with 100 U Exonuclease III (\#M0206S, NEB) to remove biotinylated free ends, followed by SPRI cleanup and elution into 101 $\mu L$ low TE

\subsection*{Shearing and Library Prep}
\label{sec-5-7}
DNA was sheared using a Diagenode Bioruptor set to 'Low' for 60 cycles of 30 seconds on / 30 seconds off.  After shearing, the DNA was filled in with Klenow polymerase and T4 PNK (\#EK0032, Thermo Scientific) at 20°C for 30 minutes.
Following the fill-in reaction, DNA was pulled down on C1 beads that had been prepared by washing twice with Tween wash buffer, before being resuspended in 200 $\mu L$ 2X NTB (2M NaCl, 10mM Tris pH 8.0, 0.1mM EDTA pH 8.0, 0.2\% Triton X-100).  Once the sample was added, the beads were incubated at room temperature for 20 minutes with rocking.  Subsequently, unbiotinylated DNA fragments were removed by washing the beads three times before resuspending in Low TE.  Sequencing libraries were generated using established protocols \citep{pmid20516186}.

\subsection*{Read mapping}
\label{sec-5-8}
Sequence reads were truncated whenever a junction was present (GATCGATC for MboI, AATTAATT for MluCI). Reads were then aligned using SMALT [\url{http://www.sanger.ac.uk/resources/software/smalt/}] with the -x option to independently align forward and reverse reads. PCR duplicates were marked using Picard-tools MarkDuplicates [\url{http://broadinstitute.github.io/picard/}]. Non-duplicate read pairs were used in analysis if both reads mapped and had a mapping quality greater than 10.

\subsection*{De novo assemblies}
\label{sec-5-9}
The human and alligator \emph{de novo} shotgun assemblies were generated with Meraculous 2.0.3 \citep{pmid21876754} using publicly available short-insert and mate-pair reads \citep{pmid22156294,pmid25504731}. The alligator mate-pair reads were adapter-trimmed with Trimmomatic \citep{pmid24695404}. Some overlapping alligator short-insert reads had been “merged.” These were unmerged back into forward and reverse reads.

\subsection*{Chicago HighRise (HiRISE) Scaffolder}
\label{sec-5-10}
\subsection*{Input pre-processing}
\label{sec-5-11}
To exclude Chicago reads that map to highly repetitive genomic regions likely to provide misleading links, we used the depth of aligned shotgun reads to identify problematic intervals.  We used a double threshold strategy: identify all intervals of the starting assembly with mapped shotgun read depth exceeding $t_1$ that contain at least one base with a mapped read depth exceeding $t_2$.  In practice we set $t_1$ and $t_2$ such that about 0.5\% of the assembly was masked. We also excluded all Chicago links falling within a 1 Kbp window on the genome which is linked to more than four other input contigs by at least two Chicago links.

\subsection*{Estimation of likelihood model parameters}
\label{sec-5-12}
Several steps of the HiRise pipeline use a likelihood model of the Chicago data to guide assembly decisions or to optimize contig order and orientation within scaffolds.  The likelihood function $L(l1,l2,g,o) = \frac{N!}{(N-n)!} (1-p_o)^{N-n} \prod_{i=1}^n f(d_i)$ gives the probability of observing the number $n$ and implied separations of spanning Chicago pairs $d_i$ betwen contigs 1 and 2, assuming the contigs have relative orientations $o \in {++,+-,-+,--}$ and are separated by a gap of length \$g.\$  The function $f(x)$ is the normalized probability distribution over genomic separation distances of Chicago read pairs, and is assumed to have a contribution from “noise” pairs  which sample the genome independently.  $f(x) = p_n/G + (1-p_n) f’(x)$, and $f’(x)$ is represented as a sum of exponential distributions.

To obtain robust estimates of $N$, $p_n$, $G$, and $f’(x)$ when the available starting assembly has limited contiguity, we first fixed an estimate of the product $N p_n$, the total number of “noise” pairs by tabulating the densities of links (defined as $n/l_1 l_2$) for a sample of contig pairs, excluding the highest and lowest 1\% of densities, and setting $N_n = G^2 \sum n_{ij} / \sum l_i l_j$, using the sum of the lengths of input contigs as the value of $G$.  We then fit the remaining parameters in $N f(x)$ by least squares to a histogram of observed separations of Chicago read pairs mapped to starting assembly contigs, after applying a multiplicative correction factor of  $G (\sum_{i=1}^{N_c} min(0,l_i-x) )^{-1}$ to the smoothed counts at separation $x$.

\subsection*{Break low-support joins in the input contigs}
\label{sec-5-13}
To identify and break candidate misjoins in the starting assembly, we used the likelihood model to compute the log likelihood change gained by joining the left and right sides of each position $i$ of each contig in the starting assembly (i.e. the log likelihood ratio (LLR) $L_i = \ln L(g=0)/L(g=\infty)$ for the two contigs that would be created by breaking at position $i$)  When this support fell below a threshold value $t_b$ over a maximal internal segment of an input contig, we defined the segment as a “low support” segment. After merging low support segments lying within 300bp of one another, and excluding those within 1 Kbp of a contig end, we either (a) introduced a break in the contig at the midpoint of the segment, or (b) if the segment is longer than 1000 bp, introduced breaks at each end of the segment.

\subsection*{Contig-contig linking graph construction}
\label{sec-5-14}
During the assembly process, the Chicago linking data was represented as a graph in which (broken) contigs of the starting assembly are nodes and edges are labeled with a list of ordered pairs of integers, each representing the positions in the two contigs of the reads from a mapped Chicago pair.  The initial steps of scaffolding were carried out in parallel on subsets of the data created by partitioning the graph into connected components by excluding edges with fewer than a threshold $t_L$ number of Chicago links, where the lowest integer threshold that did not give rise to any connected components that comprised more than 5\% of the input contigs.

\subsection*{Seed scaffold construction}
\label{sec-5-15}
The iterative phase of scaffold construction was seeded by filtering the edges of the contig-contig graph and decomposing it into high-confidence linear subgraphs.  First, the contig-contig edges were filtered and the minimum spanning forest of the filtered graph was found (see “edge filtering” below).  The graph was linearized by three successive rounds of removing nodes of degree 1 followed by removal of nodes with degree greater than 2.  Each of the connected components of the resulting graph had a linear topology and defined an ordering of a subset of the input contigs.  The final step in the creation of the initial scaffolds was to find the maximum likelihood choice of the contig orientations for each linear component.

\subsection*{Edge filtering}
\label{sec-5-16}
The following filters were applied to the edges of the contig-contig graph before linearization.   Edges from “promiscuous” contigs were excluded.  “Promiscuous” contigs were those for which the ratio of the degree in the graph of the corresponding node to the contig length in basepairs exceeds $t_p$, or have links with at least $t_L$ links  to more than $d_m$ other contigs.  The thresholds $t_p$ and $d_m$ were selected to exclude approximately 5\% of the upper tail of the distribution of the corresponding value.

\subsection*{Contig orienting}
\label{sec-5-17}
Each input scaffold can have one of two orientations in the final assembly, corresponding to the base sequences of the forward and reverse, or "Watson" and "Crick" DNA strands.  The optimal orientations for the scaffolds in each linear string was found by dynamic programming using the following recursion relationship:  In an ordered list of scaffolds of length $n$, the score of the highest-scoring sequence of orientation choices for the scaffolds up  to scaffold $i$, such that scaffolds $i-k$ to $i$ have particular  orientations $o_{i-k}, o_{i-k+1} , ... o_{i}$ is given by:

\[ S_m(i,o_{i-k},o_{i-k+1},...,o_i) = \max_{o_{i-1-k} \in [+,-]} (
S_m(i-1,o_{i-1-k},o_{i-k},...,o_{i-1}) + \sum_{j=i-i-k}^{j=i-1} \log
p(o_j,o_i) ) \]

Including links from contigs $k$ steps back provided a significant improvement in orientation accuracy because small intercalated scaffolds might only have linking and therefor orientation information on one side, with important orientation information for the flanking scaffolds coming from links that jump over it.

\subsection*{Merge scaffolds within components}
\label{sec-5-18}
Contig ends were classified as “free” if they lie at the end of a scaffold, or “buried” if they were internal to a scaffold.  For all pairs of contig ends within each connected component, the LLR score for joining them was computed with a "standard" gap size of $g_0$.  These candidate joins were sorted in decreasing order of score and evaluated according to the following criteria.  If both ends are free and from different scaffolds, test linking the two scaffolds end-to-end.  If one end is buried and the other is free, and the ends are from different scaffolds, test inserting the scaffold of the free end into the gap adjacent into the buried end.  If one or both ends is buried and the ends are on the same scaffold, test inverting the portion of the scaffold between the two ends.  If both ends are buried and from different scaffolds, test all four ways of joining the scaffolds end-to-end.  In all cases, the possible joins, insertions and inversions were tested by computing the total change in LLR score by summing the LLR scores between all pairs of contigs affected by the change.  If the change increased the LLR score, the best move was accepted.

\subsection*{Local order and orientation refinement}
\label{sec-5-19}
To refine both the local ordering and orientations of contigs in each scaffold, a dynamic programming algorithm was applied that slides a window of size $w$ across the ordered and oriented contigs of each scaffold.  At each position $i$, all the $w! 2^w$ ways of ordering and orienting the contigs within the window were considered, and a score representing the optimal ordering and orientation of all the contigs up to the end of the current window position that ends with the current O\&O of the contigs in the window was stored. The scores of all "compatible" O\&Os in windows at positions $i-1$, $i-2$, … $i-w$, and scores the extension of their orderings with the current O\&O were used.  Since $w! 2^w$ is such a steep function, the method is limited in practice to small values of $w$.

\subsection*{Iterative joining}
\label{sec-5-20}
After the initial scaffolds had been constructed within each connected component, the resulting scaffolds were returned to a single pool, and multiple rounds of end-to-end and intercalating scaffold joins were carried out.  In each round, all pairs of scaffolds were compared, and likelihood scores were computed in parallel for end-to-end and intercalating joins. The candidate joins were then sorted and non-conflicting joins were accepted in decreasing order of likelihood score increase.

\section*{Competing financial interests}
\label{sec-6}

The authors have applied for patents on technology described in this manuscript, and Dovetail Genomics LLC is established to commercialize this technology.  R.E.G. is Founder and Chief Scientific Officer of Dovetail Genomics.  D.H. and D.S.R are members of the Scientific Advisory Board.

\section*{}
\label{sec-7}
\bibliographystyle{\mybibliostyle}
\bibliography{chicago}
\end{document}